\documentclass[aps, prl, reprint, superscriptaddress]{revtex4-2}

\usepackage{amsmath}
\usepackage{newtxtext, newtxmath}
\usepackage{braket}
\usepackage{graphicx}
\usepackage{color}
\definecolor{bblue}{rgb}{0, 0.0, 0.8}
\definecolor{rred}{rgb}{0.8, 0.0, 0.0}
\definecolor{black}{rgb}{0.0, 0.0, 0.0}

\usepackage{hyperref}
\hypersetup{
    colorlinks=true,        
    linkcolor=bblue,          
    citecolor=bblue,        
    filecolor=bblue,      
    urlcolor=bblue           
}

\newcommand{\oo}{O$_{2}$\ }
\newcommand{\ooo}{O$_{2}$}

\begin{document}
\title{X-ray free-electron laser observation of giant and anisotropic magnetostriction in $\beta$-\oo at 110~Tesla}

\author{Akihiko~Ikeda}
\email[These authors contributed equally to this work: ]{a-ikeda@uec.ac.jp}
\affiliation{Department of Engineering Science, University of Electro-Communications, Chofu, Tokyo 182-8585, Japan}
\author{Yuya~Kubota}
\email[These authors contributed equally to this work: ]{kubota@spring8.or.jp}
\affiliation{RIKEN SPring-8 Center, Sayo, Hyogo 679-5148, Japan}
\author{Yuto~Ishii}
\author{Xuguang~Zhou}
\author{Shiyue~Peng}
\author{Hiroaki~Hayashi}
\author{Yasuhiro~H.~Matsuda}
\affiliation{Institute for Solid State Physics, University of Tokyo, Kashiwa, Chiba 277-8581, Japan}
\author{Kosuke~Noda}
\author{Tomoya~Tanaka}
\author{Kotomi~Shimbori}
\author{Kenta~Seki}
\author{Hideaki~Kobayashi}
\author{Dilip~Bhoi}
\affiliation{Department of Engineering Science, University of Electro-Communications, Chofu, Tokyo 182-8585, Japan}
\author{Masaki~Gen}
\affiliation{Institute for Solid State Physics, University of Tokyo, Kashiwa, Chiba 277-8581, Japan}
\affiliation{RIKEN Center for Emergent Matter Science (CEMS), Wako 351-0198, Japan}
\author{Kamini~Gautam}
\affiliation{RIKEN Center for Emergent Matter Science (CEMS), Wako 351-0198, Japan}
\author{Mitsuru~Akaki}
\affiliation{Institute for Material Research, Tohoku University, Sendai, Miyagi 980-0812, Japan}
\author{Shiro~Kawachi}
\affiliation{Graduate School of Science, University of Hyogo, Koto, Hyogo 678-1297, Japan}
\author{Shusuke~Kasamatsu}
\affiliation{Faculty of Science, Yamagata University, Kojirakawa, Yamagata 990-8560, Japan}
\author{Toshihiro~Nomura}
\affiliation{Department of Physics, Faculty of Science, Shizuoka University, Shizuoka 422-8529, Japan}
\author{Yuichi~Inubushi}
\author{Makina~Yabashi}
\affiliation{RIKEN SPring-8 Center, Sayo, Hyogo 679-5148, Japan}
\affiliation{Japan Synchrotron Radiation Research Institute (JASRI), Sayo, Hyogo 679-5198, Japan}
\date{\today}

\begin{abstract}
In strong magnetic fields beyond 100~T, the significant Zeeman energy competes with the lattice interactions, where a considerable magnetostriction is expected. However, the microscopic observation of the magnetostriction above 100~T has been hindered due to the short pulse duration of $\mu$-seconds and the coil’s destruction. Here, we report the observation of the giant and anisotropic magnetostriction of $\sim1$~\% at 110~T in the spin-controlled crystal, $\beta$-O$_{2}$, by combining the single-shot diffraction of x-ray free-electron laser (XFEL) and the newly developed portable 100~T generator (PINK-02). The very soft and anisotropic response of $\beta$-O$_{2}$ should originate in the competing van der Waals force and exchange interaction, and also the frustration of spin and lattice on the triangular network. The XFEL experiment above 100~T using PINK-02 enables microscopic investigations on materials’ properties at high magnetic fields, providing insights into how spins contribute to the stability of crystal structures.
\end{abstract}

\maketitle
Magnetic fields are fertile ground for novel findings in diverse scientific disciplines, ranging from condensed matter physics and planetary science to biological science.
The primary objective of applied magnetic fields on materials of a few to a few tens of Teslas are to {\it see} and {\it change} the electronic and magnetic states \cite{GerberScience2015, RanNatPhys2019}, which is constrained by the limited scale of Zeeman energy \footnote{\rm{Zeeman energy mounts to over 1.34 K for a free-electron spin at 1 T with the relation $E_{\rm{Zeeman}} = g\mu_{\rm{B}}S_{z}B_{z}$. $g$, $\mu_{\rm{B}}$, $S_{z}$, and $B_{z}$ are $g$-factor, Bohr magneton, Spin, and external magnetic field, respectively.}}.
On the other hand, magnetic fields beyond $\sim100-1000$ T are the frontier where they are expected to {\it break} the crystal state of materials.
Zeeman energy at 1000~T amounts to 1340~K, which competes the crystal lattice's energy.
Even with a magnetostriction of $\Delta L/L = 10^{-6}$ at 1~T, the extrapolation with the relation $\Delta L/L \propto B^{2}$ gives a significant magnetostriction of 1~\% and 100~\% at 100 and 1000~T, respectively, making us believe that some drastic change in the relation between spin and lattice happens above 100~T.
Meanwhile, the magnetic field can also be a valuable tool for modifying the crystal in a manner distinct from other stimuli, such as pressure, strain, chemical substitution, and photo-irradiation. 

\begin{figure*}[t]
\begin{center}
\includegraphics[width = \textwidth]{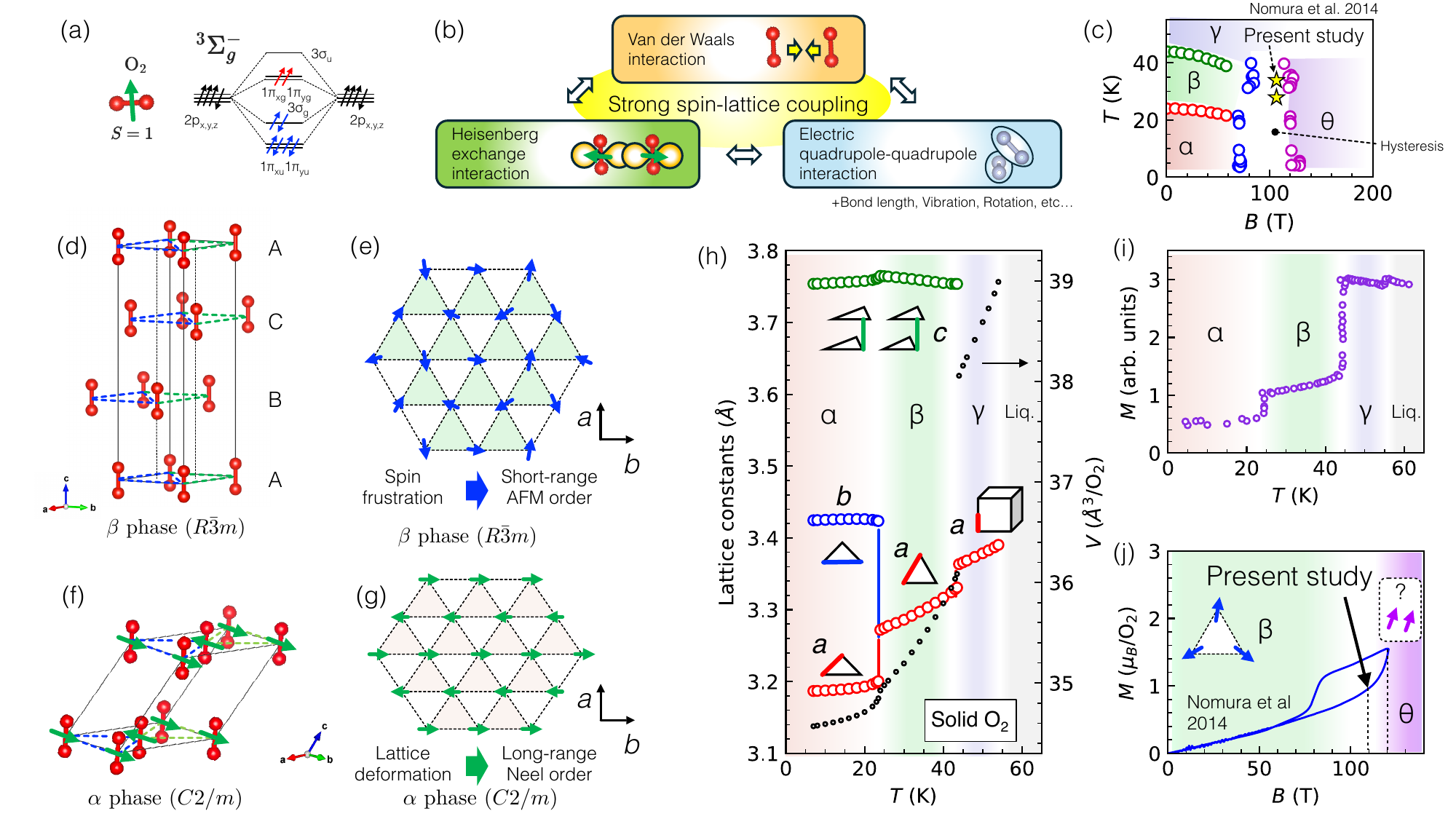}
\caption{
(a) The molecular orbital of \oo with $S=1$. 
(b) The competing interactions in solid \ooo. 
(c) A phase diagram of solid \oo on a temperature-magnetic field plane reproduced from Ref. \cite{NomuraPRB2017}.
(d) The crystal structure of $\beta$-\oo whose space group is $R\bar{3}m$.
Layers of a triangular lattice are stacked vertically in the ABC-ABC manner. 
(e) The schematic of spin configuration on the triangular lattice of $\beta$-\ooo.
Due to an intense frustration from the lattice symmetry, the spin shows a short-range order with a 140$^{\circ}$ structure \cite{DunstetterJMMM1988}.
(f) The crystal structure of $\alpha$-\oo whose space group is $C2/m$.
It is a deformed structure of $\beta$-\ooo.
In the layer, the triangle lattice is deformed so that the spin frustration is relaxed.
(g) The schematics for the arrangement of the N\'{e}el order of spins on the deformed triangular lattice in $\alpha$-\ooo, which results in the non-equivalent exchange constant in $a$ and $b$ direction.
(h) The temperature dependence of lattice parameters and volume in solid \ooo.
The data are extracted from Ref. \cite{FreimanPhysRep2004}.
(i) The temperature dependence of magnetization of solid \oo.
The data is extracted from Ref. \cite{MeierJPC1982}.  
(j) The magnetization of solid \oo at 32~K up to 124~T. 
The data is adopted from Ref. \cite{NomuraPRB2015}.
Magnetic field induced phase transition from $\beta$-\oo to $\theta$-\oo occur above 120~T.
In the present study, we generated up to 110~T, where we obtained XRD of the strained $\beta$-\ooo by magnetostriction.
\label{intro}}
\end{center}
\end{figure*}

Solid O$_{2}$, a spin-controlled crystal, is the representative candidate for a magnetic field-induced lattice change where the strong spin-lattice coupling is in play \cite{FreimanPhysRep2004, FreimanPhysRep2018}.
\oo is a molecular magnet with $S=1$ with molecular orbital state $^{3}\sum^{-}_{g}$ as depicted in Fig. \ref{intro}(a).
Upon cooling, it condenses into a liquid, and then three distinct solids, $\gamma$-\ooo, $\beta$-\ooo, and $\alpha$-\ooo.
The variety of the solid phases is the manifestation of the competing interactions in solid O$_{2}$ such as van der Waals interaction, the interaction between the electric quadrupole moments (These two factors breeds the exotic phase diagram of solid N$_{2}$ \cite{KirszPRB2024}), and the antiferromagnetic (AFM) exchange interactions between the molecular spins as schematically shown in Fig. \ref{intro}(b).
The competing nature in solid \oo is even more evident in extreme conditions \cite{FreimanPhysRep2018}.
Under high pressure, three more phases show up, where the metallic phase emerges beyond 100 GPa, which becomes superconducting below 0.6 K \cite{ShimizuNature1998}.
Above an ultrahigh magnetic field of 120~T, the ferromagnetic solid \ooo, $\theta$-\ooo, was discovered a decade ago as shown in Fig. \ref{intro}(c) and \ref{intro}(j) \cite{NomuraPRL2014, NomuraPRB2015, NomuraPRB2017, NomuraPRB2017b, NomuraOxygen2022}.

We work on $\beta$-\oo, which exhibits softness and fluctuations in lattice and magnetism.
The crystal structure of $\beta$-\oo is represented by the triangular lattice in the $ab$ plane, which is stacked in the $c$ axis. 
The molecular axis aligns perpendicular to the $ab$ plane to enhancing the exchange interaction between molecular spins by maximizing the overlap of the $\pi$ orbitals as schematically shown in Fig. \ref{intro}(d).
The spin interaction in $\beta$-\oo is represented by the Heisenberg model with $S=1$ on the triangular lattice, which faces the geometrical frustration resulting in the suppression of the long-range order and realization of the short-range antiferromagnetic order of 140$^{\circ}$ structure \cite{DunstetterJMMM1988} as shown in Fig. \ref{intro}(e).
Below 24~K, the lattice exhibits a first-order structural transition to $\alpha$-\oo with monoclinic structure as shown in Fig. \ref{intro}(f) where the triangle lattice becomes significantly deformed in the $ab$ plane.
This relaxes the geometrical frustration of the spin system, resulting in the long-range N\'{e}el order as shown in Fig. \ref{intro}(g).
The deformation is evident in the lattice constant in Fig. \ref{intro}(h), showing the splitting of the length of the sides of the triangle by 7~\% upon entering the $\alpha$ phase.
Meanwhile, the magnetization shows a sudden decrease upon entering the $\alpha$ phase from the $\beta$ phase, as shown in Fig. \ref{intro}(i).
Surprisingly, the lattice parameter change of the $ab$ plane in $\beta$-\oo amounts to 2~\% in the limited temperature range from 24 to 44 K.
In contrast, comparably small changes  $<0.5$~\% are observed in the $c$ axis in $\beta$-\oo in the same temperature range.
This is due to the soft triangular lattice, which is realized by the geometrical frustration of the spin system on the triangular lattice unique to the $ab$ plane of $\beta$-\oo with the strong spin-lattice coupling \cite{BarylnikLTP1994}.

\begin{figure*}[t]
\begin{center}
\includegraphics[width = \textwidth]{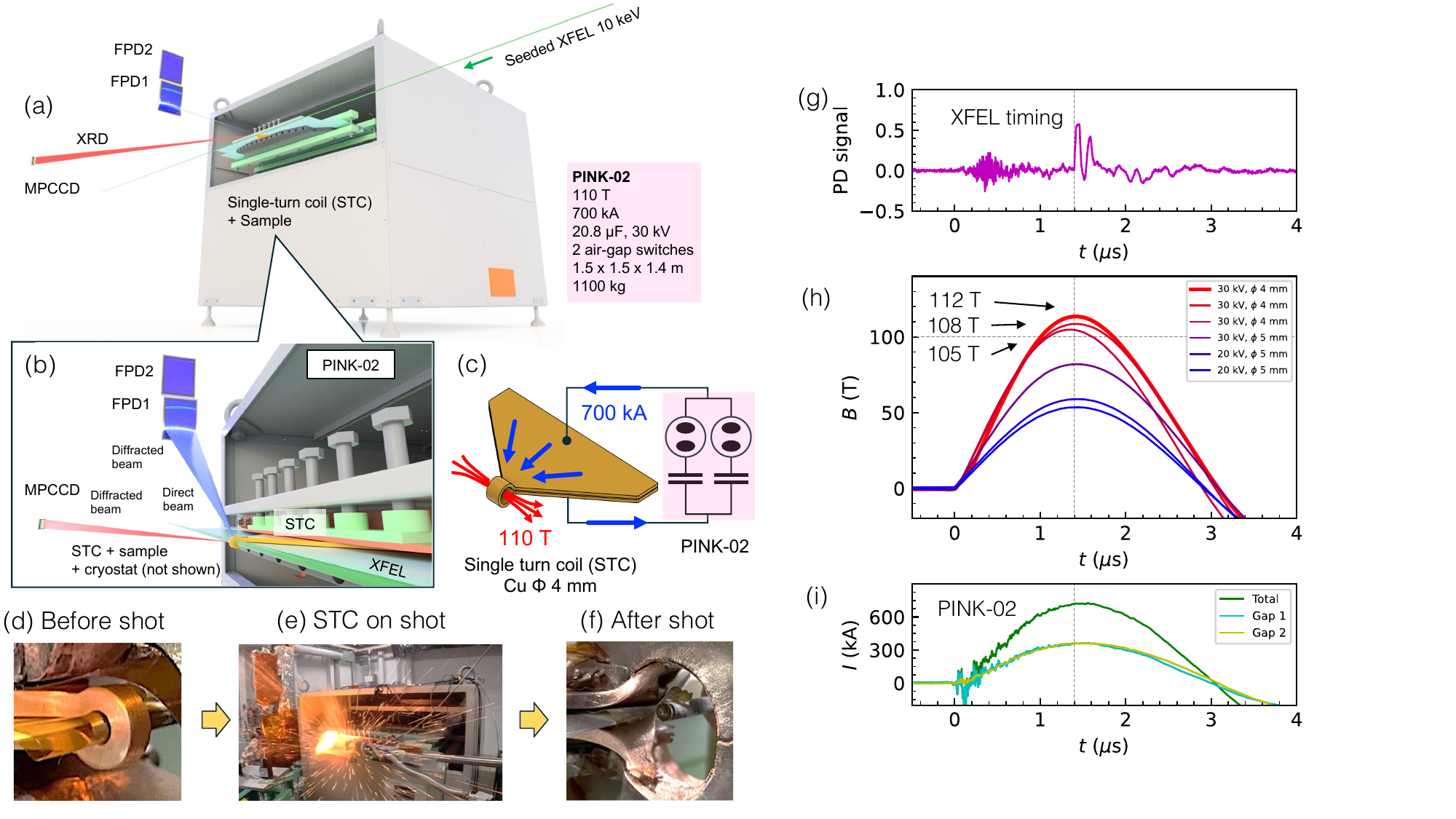}
\caption{
(a) A schematic view of PINK-02 and the arrangement for the x-ray diffraction in combination with the XFEL installed in SACLA.
(b) A magnification at the single-turn coil (STC) and x-ray beam incident on the sample, with diffracted beams propagating to the 2D x-ray detectors.
(c) A schematic drawing of the electric circuit of PINK-02 and a single-turn coil.
(d) A photo of a single-turn coil of $\phi4$ mm before the shot.
(e) A photo of the single-turn coil during the magnetic field generation.
(f) A photo of a single-turn coil after the shot.
(g) X-ray photodiode (PD) signal intensity as a function of time.
(h) Representative waveforms of pulsed magnetic field generated using PINK-02 with single-turn coils of $\phi4$ and 5 mm diameter, and charging voltages of 20 and 30 kV.
(i) Representative waveforms of the current injected into the single-turn coil.
There are two switches connected in parallel.
\label{pink02}
}
\end{center}
\end{figure*}

When ultrahigh magnetic fields beyond 100~T are applied to that spin-lattice coupled system, we expect drastic changes in the crystal parameters.
Indeed, the ferromagnetic $\theta$-\oo is discovered above 120~T as shown in Fig. \ref{intro}(j) \cite{NomuraPRL2014, NomuraPRB2015, NomuraPRB2017, NomuraPRB2017b, NomuraOxygen2022}, where the determination of crystal structure change has remained elusive.
For this purpose, we have realized a method for the macroscopic measurement of $\Delta L/L$ beyond 100~T employing the fiber Bragg grating (FBG) method \cite{IkedaRSI2017}, which has been applied to various systems \cite{IkedaJPSJ2024, IkedaPRL2020, NomuraPRB2021, IkedaNC2023, NomuraNC2023, GenPNAS2023} up to 600~T generated using single-turn coil method and the electro-magnetic flux compression method \cite{NakamuraRSI2018}.
To gain further insights into the spin-lattice coupling beyond 100~T, one needs the microscopic probes, such as x-ray diffraction (XRD), in addition to the FBG dilatometry.
However, x-ray diffractometry at 100~T is experimentally challenging because experimentally obtaining 100~T necessitates destructive pulse magnets \cite{HerlachRPP1999}.
To overcome the difficulty, we reported an XRD study with a generation of 77~T pulsed magnetic field by implementing the first Portable INtense Kyokugenjiba (PINK-01) in combination with the single-shot powder diffraction of x-ray free-electron laser (XFEL) from solid sample \cite{IkedaAPL2022, IkedaPRR2020}.
Although the $\mu$-second pulse of 77~T generated using destructive magnets was shown to be compatible with the XFEL pulses, the field strength of PINK-01 is not sufficient for the exploration beyond 100~T.

Here, we present the portable generation of 110~T using the newly devised PINK-02, which goes with the single-shot XRD for $\beta$-\ooo.
The setup is depicted in Figs. \ref{pink02}(a) and \ref{pink02}(b). 
PINK-02 employs two capacitors of 20.8 $\mu$F in total rated at 30 kV, equipped with a discharge air gap switches as schematically depicted in Fig. \ref{pink02}(c), which weighs 1100 kg. 
The capacitance and rated current of PINK-02 are twice those of PINK-01.
By injecting the current into the single-turn coil, one can obtain the ultrahigh magnetic field, where the single-turn coil explodes at each shot, as shown in Figs. \ref{pink02}(d)-(f) and Endmatter.
The average magnetic field obtained was $110 \pm 5$ T with the inner diameter of 4 mm as schematically shown in Fig. \ref{pink02}(h), where the maximum current and voltage are about 700 kA and 30 kV, respectively, as shown in Fig. \ref{pink02}(i).
The pulse duration of the magnetic field was 3 $\mu$s, whose maximum timing is well synchronized with the XFEL pulse as shown in Fig. \ref{pink02}(g).
The experiment was carried out at BL3 in SACLA \cite{IshikawaNP2012}.
XFEL pulses at 10 keV with self-seeding \cite{InoueNP2019} with a dispersion of $1\sim3$ eV is delivered to the powder sample with $\sim360$~$\mu$J/pulse.

\begin{figure*}
\begin{center}
\includegraphics[width = \textwidth]{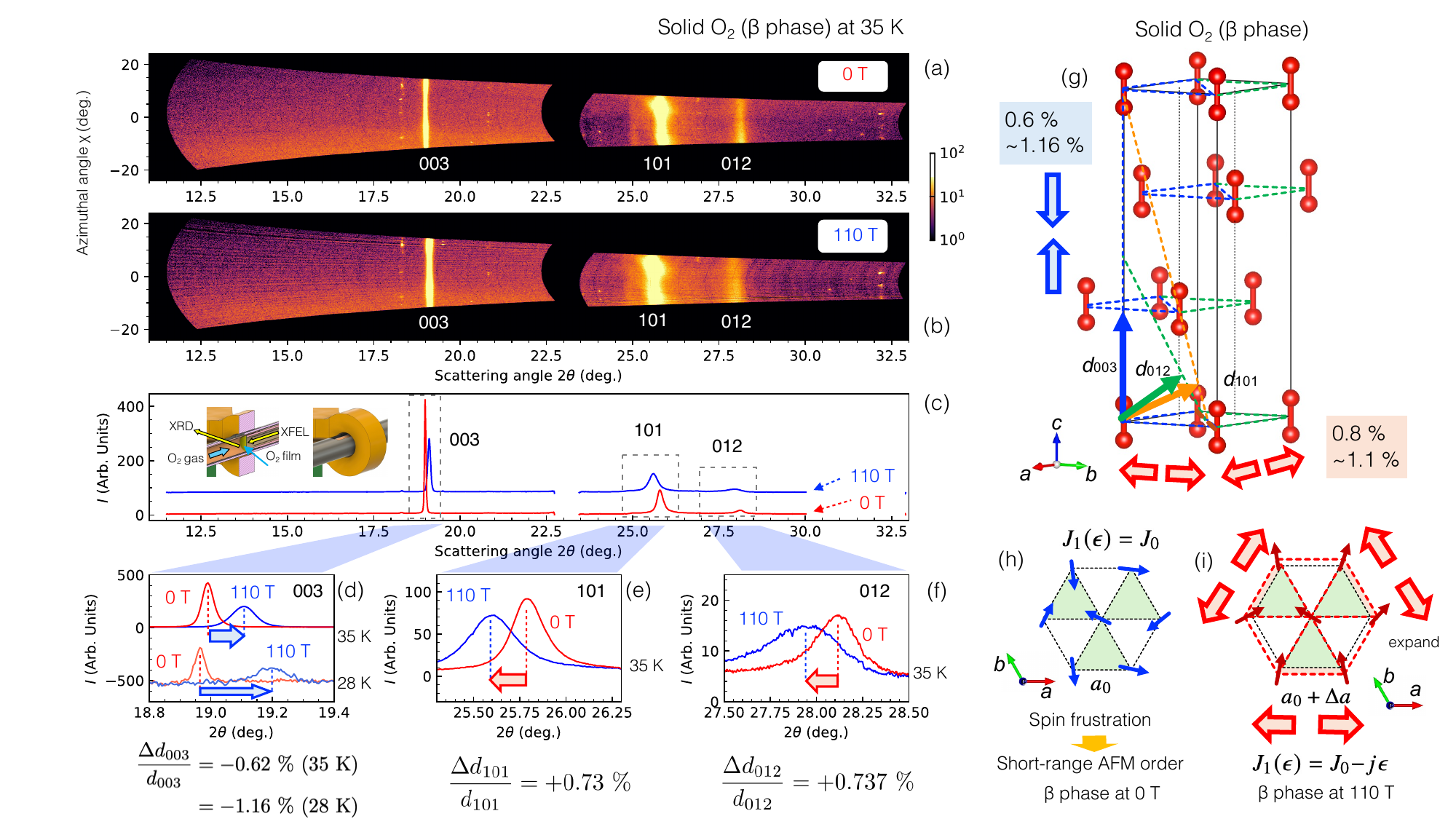}
\caption{(a) Powder XDR data from $\beta$-\oo recorded using two-dimensional x-ray detectors at 0~T and
(b) at 110~T.
(c) Integrated powder XDR data from $\beta$-\oo at 0~T and at 110~T.
The inset shows the schematic drawing of the experimental set up at the sample position.
(d) The magnified powder XRD data at the 003 reflection of $\beta$-\ooo at 35 and 28~K, showing -0.62 and -1.16~\% shrinkage of the inter-plane separation at 110~T, respectively.
(e) The magnified data of powder XRD at the 101 reflection of $\beta$-\ooo at 35~K, showing +0.73~\% elongation of the inter-plane separation at 110~T.
(f) The magnified data of powder XRD at the 012 reflection of $\beta$-\ooo at 35~K, showing +0.737~\% elongation of the inter-plane separation at 110~T.
(g) The schematic drawing of the crystal structure of $\beta$-\oo with the indication of the shrinkage in the $c$ axis and elongation in the $ab$ plane.
The diffraction planes are also depicted.
(h) The schematic drawing of the crystal structure and spin configuration in $\beta$-\oo at 0~T, where the \oo molecules form the triangular lattice with the short-range order due to the antiferromagnetic coupling and geometrical frustration.
(i) The schematic drawing of the crystal structure and spin configuration in $\beta$-\oo at 110~T, where the spins are more aligned to the external magnetic field, losing the antiferromagnetic short-range order.
This results in the expansion of the lattice parameter in $ab$ plane to relax the antiferromagnetic exchange coupling between spins.
 \label{result}}
\end{center}
\end{figure*}

The results of the powder XRD of $\beta$-\oo obtained at 0~T and 110~T are displayed in the full image of the x-ray detector in Figs. \ref{result}(a) and \ref{result}(b), respectively.
They show three XRD peaks from solid \ooo, 003, 101, 012, by which the solid \oo is attributed to be in $\beta$-\oo.
By comparing the data in Figs. \ref{result}(a) and \ref{result}(b), one can tell that the peak positions are shifted in the data at 110~T.
The shift is qualitatively shown in the plots in Figs. \ref{result}(c)-(f), where the signs of the shifts are dependent on the diffraction indexes.
In the present experimental geometry, the observed lattice strain is transverse primarily rather than longitudinal magnetostriction.
Although we obtained only two data sets in the present study, the reliability of our measurement is ensured by several other samples, including ${\mathrm{CuGaCr}}_{4}{\mathrm{S}}_{8}$ \cite{GenPRB2025}.

The anisotropic magnetostriction is evidenced by observing the positive shift of the 003 diffraction angle, and the negative shift of the  101 and 012 diffraction angles, respectively, at 110~T.
In Fig. \ref{result}(d), the 003 diffraction peak at 28~K is shown along with that at 35~K, where the scattering angles $2\theta$ show shifts from 18.99$^{\circ}$ to 19.11$^{\circ}$ and from 18.965$^{\circ}$ to 19.19$^{\circ}$, respectively.
With the definition $\Delta d_{hkl}/d_{hkl} \equiv \left(d_{hkl}^{\rm{110\ T}}/d_{hkl}^{\rm{0\ T}}\right) - 1 = \left(\sin\theta_{hkl}^{\rm{0\ T}} / \sin\theta_{hkl}^{\rm{110\ T}}\right) - 1$ \footnote{\rm{Deduced from the following relation, $\lambda = 2d^{\mathrm{0\ T}}_{hkl}\sin\left(\theta^{\mathrm{0\ T}}_{hkl}\right) = 2d^{\mathrm{110\ T}}_{hkl}\sin\left(\theta^{\mathrm{110\ T}}_{hkl}\right)$.}}, the values of $\Delta d_{003} / d_{003} = -0.62$~\% and -1.16~\% are deduced, respectively, which readily correspond to the shrinkage of the $c$ axis ($\Delta c/c<0$) as shown in Fig. \ref{result}(g).
These are twice the value of the thermally induced change of the $c$ axis in the $\beta$ phase as shown in Fig. \ref{intro}(i), which is only 0.5~\%.

In Figs. \ref{result}(e) and \ref{result}(f), for the 101 and 012 diffraction peaks, $2\theta$ show shifts from 25.79$^{\circ}$ to 25.6$^{\circ}$ and  from 28.12$^{\circ}$ to 27.91$^{\circ}$, respectively.
The deduced values are $\Delta d_{101} / d_{101} = +0.73$~\% and $\Delta d_{012} / d_{012} = +0.737$~\%, respectively.
With the pairs of the values $\Delta d_{hkl} / d_{hkl}$ ($hkl = 003$ and 101, or 003 and 012), one can deduce the values of $\Delta a/a = 0.8$, or 1.1~\%, respectively.
The reason for the discrepancy in the values is unclear.
With an averaged value $\Delta a/a = 0.95$~\% and $\Delta c/c = -0.62$~\% at 110~T and 35~K, one can deduce the following the values, the volume expansion $\Delta V/V\equiv \Delta c/c + \Delta a/a \times 2 = 1.28$~\% and the distortion value $\Delta D/D \equiv  \Delta c/c - \Delta a/a = -1.57$~\%.
Note that $\Delta D/D = 0$ if the lattice change is isotropic (i.e. $\Delta a/a = \Delta c/c$).
The anisotropic and significant magnetostriction is indicated by the obtained values of $\Delta a/a = 0.95$~\%, $\Delta c/c = -0.62$~\%, $\Delta V/V = 1.28$~\%, and $\Delta D/D = -1.57$~\% in the $\beta$ phase of solid \oo at 110~T and 35~K.
An even more significant value of $\Delta c/c = -1.16$~\%  at 110~T and 28~K is noteworthy.

We discuss the origin of the observed anisotropic magnetostriction. 
$\beta$-\oo is regarded as a two-dimensional spin system, where the exchange interaction is strong within the $ab$ plane and subtle between the planes \cite{FreimanPhysRep2004}.
To discuss this in detail, we consider a two-dimensional Heisenberg model with exchange striction, as follows \cite{IkedaPRB2019}, $\mathcal{H} = \sum_{\braket{i, j}} J_1(\mathbf{R}_{i,j}) \hat{\mathbf{S}}_{i} \cdot \hat{\mathbf{S}}_{j} + k\mathbf{\epsilon}^2/2 - g\mu_{\rm{B}}\sum_{i} \hat{\mathbf{S}}_{i} \cdot \mathbf{B} $, where $J_1(\mathbf{R}_{i,j})$, $\braket{i, j}$, $\hat{\mathbf{S}}_{i}$, $k$, $\epsilon$, $g$, $\mu_{\rm{B}}$, and $\mathbf{B}$ are exchange constant dependent on intermolecular arrangement $\mathbf{R}_{i,j}$, pairs of the nearest neighbor sites, spin operator, the elastic constant, and the lattice strain,  $g$-factor, Bohr magneton, and magnetic field.
The first, the second, and the third terms of the Hamiltonian describe the antiferromagnetic exchange coupling between spins on the nearest neighboring sites of the triangle lattice, the elastic energy of the lattice in the $ab$ plane, and the Zeeman energy for individual spins, respectively.
For simplicity, we consider that $J_1(\mathbf{R}_{i,j})$ depends on $\epsilon$.
We obtain ${J_1(\epsilon)} = J_{0} - j \epsilon$, where $J_{0}$ and $j$ are the exchange constant without strain and the coefficient for the strain-dependent change of exchange constant.
The Hamiltonian explains the lattice expansion in the $ab$ plane.
First, the magnetization gradually increases up to 0.9~$\mu_{\rm{B}}$/\oo \cite{NomuraPRB2015} at 110~T, where the Heisenberg term of the Hamiltonian increases the total energy. 
The system attempts to compensate for the increased energy by minimizing the exchange constant $J_1(\epsilon)$, which is achieved by increasing the intermolecular distance in the $ab$ plane with a decrease of the energy of $j\epsilon \braket{\hat{\mathbf{S}}_{i} \cdot \hat{\mathbf{S}}_{j}}$.
$\epsilon$ increases until the Heisenberg energy is equilibrated with the rise of the elastic energy $k\epsilon^2 / 2$.

\begin{figure}
\begin{center}
\includegraphics[width = \columnwidth]{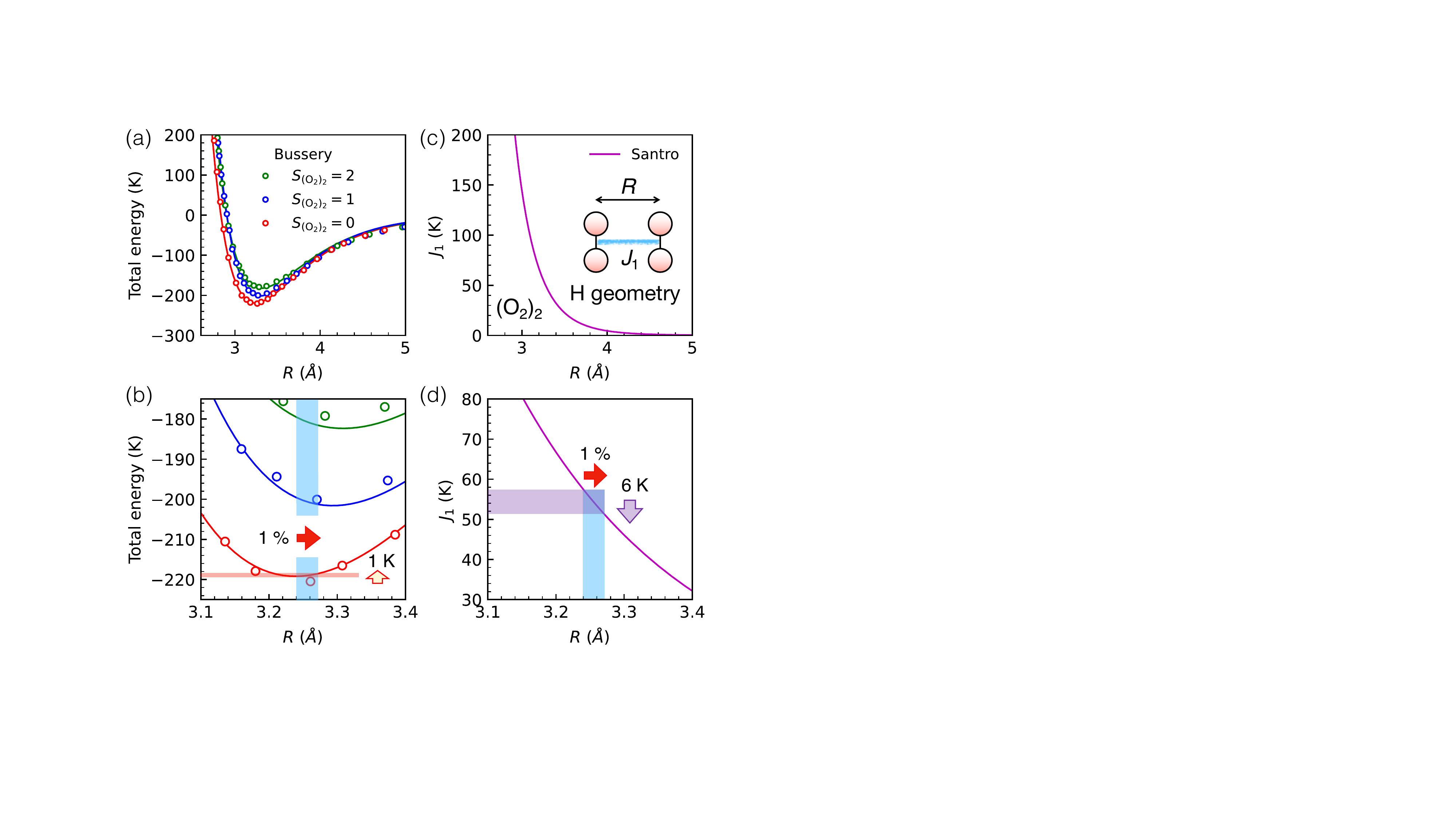}
\caption{
(a) Intermolecular potential of the dimer (\ooo)$_{2}$ calculated as a function of the inter-molecular distance $R$  in H geometry with a variation of the total spin $S_{(\rm{O}_{2})_{2}} = 0$, 1, and 2 \cite{BusseryPRB1993}.
(b) A magnification of (a).
(c) Exchange constant $J_1$ of between molecular spins as a function of $R$ deduced from the high-pressure experiment \cite{SantoroPRB2001}.
(d) A magnification of (c).
\label{pot}}
\end{center}
\end{figure}

To gain a qualitative insights into the balance of the elastic energy and the exchange energy, we refer to the spin-dependent intermolecular potential calculated for the dimer (\ooo)$_{2}$  in the parallel (H) geometry \cite{BusseryPRB1993} and the spin-spin exchange constant $J_1$ estimated in a high-pressure experiment \cite{SantoroPRB2001}, which shows a qualitative agreement with our discussion.
Figs. \ref{pot}(a) and \ref{pot}(b) shows the intermolecular potential as a function of the inter-molecular distance $R$ with a variation of $S_{(\rm{O}_{2})_{2}} = 0$, 1, and 2 \cite{BusseryPRB1993}, which is fitted to the Morse potential \footnote{$M(R) = d\left(1- \left(1 - \exp\left(- \left(R - R_{0}\right)/a\right)\right) ^{2}\right)$ with $d$, $a$, and $R_{0}$ being the depth, the inverted width, and the equilibrium position of the potential, respectively}.
As shown in Fig. \ref{pot}(b), with increasing $R$ by 1~\% from the equilibrium position, the increase of the elastic energy ($k\epsilon^2 / 2$) is $\sim1$ K/bond, which amounts to +3 K/\ooo.
Further, the dependence of $J_1$ on the intermolecular distance of \oo is estimated in a high-pressure study of solid \oo \cite{SantoroPRB2001} as shown in Figs. \ref{pot}(c) and \ref{pot}(d). 
With increasing $R$ by 1~\%, the decrease of $J_1$ ($=j\epsilon$) is $\sim-6$ K/bond.
With the magnetization value at 110~T of $g\mu_{\mathrm{B}}\braket{S^z} =$ 0.9 $\mu_{\mathrm{B}}$ \cite{NomuraPRB2015} and assuming $j\epsilon \braket{\hat{\mathbf{S}}_{i} \cdot \hat{\mathbf{S}}_{j}}\simeq  j \epsilon \braket{\hat{{S}}^z_i}\braket{\hat{S}^z_j}$, the reduction of the exchange energy amounts to an $\sim-3.6$ K/\ooo.
One finds that it is a comparable value for the aforementioned increase of the elastic energy.

The qualitative argument above with the dimer (O$_2$)$_2$ is further strengthened by our first-principles calculation of the exchange constants in the solid $\beta$ phase with the structure of 0~T and the deformed unit cell at 110~T obtained in the present study.
Here, we denote the nearest-neighbor exchange interaction in the $ab$ plane direction and in the interlayer ($c$) direction as $J_1$ and $J_2$, respectively. 
The calculation employs the vdW density functional theory and the tight-binding model calculation.
The calculated result of $J_{1}$(0~T) and $J_{1}$(110~T) are 47.35 and 41.8 K/bond, where $\Delta J_{1} = -5.55$ K is an 11.8~\% reduction of the AFM interaction, being in good agreement with the reported results on the dimer (\ooo)$_{2}$ as shown in Fig. \ref{pot}(d) and the exchange striction mechanism driving the expantion in the $ab$ plane.
The calculated result of $J_{2}$(0~T) and $J_{2}$(110~T) are 6.5 and 7.4 K/bond, where $\Delta J_{2} = +0.9$ K is an 14.3~\% increase of the AFM interaction.
$\Delta J_{2}$ is less significant than $\Delta J_{1}$ and the sign counters to the observed lattice change.
This indicates that the shrinkage in $c$ axis at 110~T is not driven by the exchange striction mechanism with $\Delta J_{2}$, which accords with the previous notion that the inter-layer coupling of spins is negligible due to the small hybridization of the $\pi$ orbitals  \cite{BusseryJCP1994}.
Another possible mechanism for the shrinkage in $c$ axis is the conservation effect of the lattice volume expansion with the expanding $ab$ plane, which is not conclusive.

In conclusion, we obtained powder x-ray diffraction data of $\beta$-\oo\ at 110~T by establishing the novel portable 100~T technique PINK-02 compatible with XFEL.
An anisotropic lattice deformation of up to 1~\% was observed, which arises from the two-dimensionality of the spin system and the competition between exchange and van der Waals interactions.
Our method opens the way to exploring spin–lattice coupling phenomena above 100~T.
The clarification of the crystal structure of $\theta$-\oo is, without a doubt, the most crucial goal of our study to follow. 

\begin{acknowledgements}
Fruitful discussion with J. Nasu is acknowledged.
The experiment was conducted with the approval of JASRI (Proposal Nos. 2024A8010 and 2024B8046).
The authors would like to acknowledge the support from the technical staff of the SACLA facility.
PINK-02 is established with the support of the SACLA/SPring-8 Basic Development Program (2021 - 2024).
This work is supported by the JST FOREST (Program Nos. JPMJFR222W and JPMJFR2037), JSPS Grant-in-Aid for Scientific Research on Innovative Areas (A) (1000~T Science) 23H04861, 23H04859, 24H01633, Grant-in-Aid for Scientific Research (B) 23H01121, Grant-in-Aid for Scientific Research 24K21043, and MEXT LEADER program No. JPMXS0320210021.
\end{acknowledgements}

\paragraph{Data Availability}
The data that support the findings of this article are openly available \footnote{\rm{https://doi.org/10.5281/zenodo.17166637}}.

\bibliography{pink02}

\clearpage

\section{Endmatter}

\begin{figure}
\begin{center}
\includegraphics[width = \columnwidth]{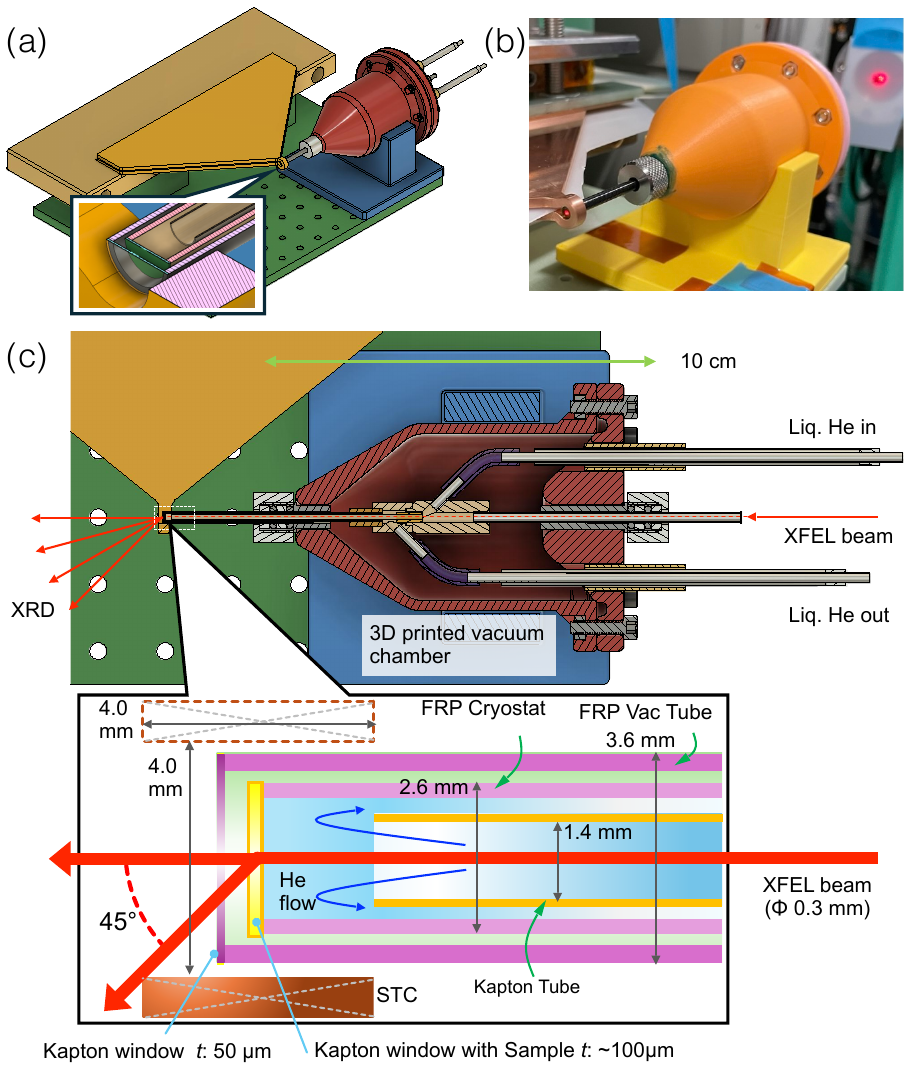}
\caption{
(a) A schematic of the micro-cryostat, 3D printed vacuum chamber, and the single-turn coil setup.
(b) 
A photo of the micro-cryostat, the single-turn coil, and the 3D printed vacuum chamber.
(c) Detailed cross-sectional view of the micro-cryostat, 3D printed vacuum chamber, and the single-turn coil.
\label{cryo}}
\end{center}
\end{figure}

\paragraph{Micro-cryostat}
The all-plastic He-flow micro-cryostat is newly devised for cooling the sample at the coil center down to 10 K, allowing the incident x-ray beam of $\phi\sim0.3$ mm to hit the sample, whose overview and a photo are shown in Figs. \ref{cryo}(a) and \ref{cryo}(b).
As shown in the detailed view in Fig. \ref{cryo}(c), the diffracted beam exits from the cryostat, allowing a wide range of scattering angle $2\theta$ from 0$^{\circ}$ to 45$^{\circ}$.
The outer diameter of the micro-cryostat is $\phi$2.6 mm, which is in the vacuum tube of $\phi$3.6 mm at the center of the single-turn coil whose inner diameter is $\phi$4 mm, as schematically depicted in Fig. \ref{cryo}(c).
Micro-cyostat is made of fiber-reinforced plastic and the glue SK-229 (Nitto).
The vacuum tube and the micro-cryostat are held by the 3D printed vacuum chamber made of PLA plastic, whose inner surface is coated with the glue SK-229.

\paragraph{Preparation of solid \ooo}
The film of solid \oo is condensed at the center of single-turn coil.
As shown in the inset of Fig. 3(c), our micro-cryostat is further modified for the \oo sample, where the \oo gas inlet is directly connected to the cold-head of the cryostat.
The vacuum tube is also elongated to the beam downstream, with a hole covered with Kapton film ($t=50$ $\mu$m) so that the diffracted x-ray comes out.
First, we keep the \oo gas inlet in a vacuum while we cool the micro-cryostat.
Next, we introduce \oo gas for a limited amount, which is condensed onto the cold surface of the micro-cryostat, and then we evacuate the \oo gas inlet to the vacuum again.
We can observe the XRD from the film sample of solid \oo.
We estimate the sample temperature from the XRD to be 28 or 35~K.
We could not obtain a lower temperature than those, although 10 K is achieved with pre-prepared solid-state samples.
It may be because the additionally installed \oo gas line works as an additional heat source.

\paragraph{X-ray detectors}
The diffracted beam is detected using two kinds of 2D x-ray detectors. 
The other is a pair of flat panel detectors (FPD) with a distance of 1.0 m from the sample.
One is the multi-port charge-coupled device (MPCCD) \cite{KameshimaRSI2014} with a distance of 1.5 m.
Single-shot data are recorded at 10 Hz.
Three x-ray images are recorded sequentially at $t=-0.1$, 0, and 0.1 ms, where $t=0$ ms is at the maximum magnetic field.
The data of MPCCD are not presented in the present Letter.

\paragraph{Estimation of exchange interaction using calculations}
The first-principles calculation of the exchange constants between the
molecular spins of O$_2$ in solid O$_2$ was performed using VASP \cite{kresse1996} in combination with Wannier90 \cite{pizzi2020} and TB2J \cite{He2021} packages.
First, VASP is used to obtain the electronic structure of $\beta$ phase oxygen using the vdW-DF-optB86b functional \cite{Klimes2011} in combination with a Hubbard-$U$ correction \cite{Dudarev1998} of $U_\text{eff} = 12$ eV, which resulted in a good description of the $\alpha$ phase of solid oxygen in a previous work \cite{Kasamatsu2017}.
The lattice constants were fixed to those determined experimentally in this work, the O--O bond length was fixed to 1.207 \r{A}, and the spin configuration was fixed to the ferromagnetic one. The unit cell in the monoclinic setting was employed with a $5 \times 6 \times 5$ $k$-point mesh and a plane wave cutoff energy of 1500 eV.  
Wannier90 was used to obtain Wannier functions and build a tight-binding model with initial projections on the $p_x$ and $p_y$ orbitals of O atoms for $\pi$ and $\pi^\ast$ bands and $s$ orbitals placed at O-O bond centers for the $\sigma$ bonding band.
Only the $\pi$ and $\pi^\ast$ bands contribute to the spin interactions.
The resulting tight-binding model was fed into TB2J code, which uses the Liechtenstein formalism to calculate Heisenberg exchange constants.

\paragraph{Qualitative comparison of the lattice change  with a previous XRD report}
The significance of the lattice change is consistent with previous XRD studies on the $\alpha$ phase of solid \oo at 5~T \cite{KatsumataJPCM2005} and 28~T \cite{MatsudaPRB2019}.
First, a large $\Delta V/V = 1$~\% is reported at 5~T at 7~K \cite{KatsumataJPCM2005}, later confirmed as an extrinsic effect by Ref.  \cite{MatsudaPRB2019}.
Ref.  \cite{MatsudaPRB2019} further put a constrain that $\Delta d/d < 1\times 10^{-4}$ at 5~T and $\Delta d/d < 2\times 10^{-3}$ at 25~T.
If we assume the quadratic dependence $\Delta d/d \simeq pB^{2}$ with typical values of $\Delta d/d = 1 $~\% and $B = 100$~T, we obtain $\Delta d/d\simeq 2.5\times10^{-5}$ at 5~T, $6.3\times10^{-4}$ at 25~T by an interpolation, which are consistent with the previous XRD study up to 28~T \cite{MatsudaPRB2019} and the present result.

\paragraph{Cooperative softness of the lattice and spin system in $\beta$-\oo}
We note that the considerable fluctuation of both the spins on the triangular lattice and the arrangement of the molecular axes of $\beta$-\oo may assist the observed significant magnetostriction.
The frustrated molecular axis is inferred by the fact that no phase in solid N$_2$ shows a similar colinear configuration of the molecular axes \cite{KirszPRB2024} and also by the thermally-induced inverse broadening effect of the XRD peak in $\beta$-\oo \cite{BarylnikLTP1994}.
The significant temperature-induced change of the lattice within the $\beta$ phase of \oo (Fig. \ref{intro}(h)) is unique in contrast to the less significant lattice change in $\alpha$-\oo, where both the spin and lattice are more rigid with the N\'{e}el order.

\paragraph{Analysis of the XRD peak profile}
The XRD peak widths exhibit a clear dependence on $2\theta$, and all peaks are broader at 110~T, as shown in Figs. \ref{result}(d)-\ref{result}(f) (See \footnote{See the Supplemental Material for the qualitative analysis of the XRD peak profile.}).
Possible origins are micro-strain effect due to anisotropic magnetostriction in random powder, crystal size effect \cite{ZhaoJAC2008, MurrietaMTP2022}, and the precursor of the phase transition to $\theta$-\oo \cite{NomuraPRL2014} above 120~T; however, they need further evaluation.

\paragraph{Comparison with other giant magnetostriction systems}
The present lattice change occurs in the unit cell without a phase transition, which is considered the greatest ever reported to our best knowledge.
Larger magnetostrictions are observed in Heusller alloy \cite{KakeshitaMRS2002} and manganite systems \cite{TokuraRPP2006}, where the domain's reorientation and the phase transition are the sources of the significant lattice change.

\paragraph{Software}
The obtained image is analyzed using a python package pyFAI  \cite{KiefferJPCS2013}. The crystal structures are rendered using the 3D visualization tool VESTA \cite{MommaJAC2011}.

\paragraph{Video of destructive generation of 110~T using a portable 100~T generator PINK-02}
See the video of PINK-02 generating 110~T at BL3 in SACLA 13 times successfully in a row (37 seconds) \footnote{See the Supplemental Video for the portable 110 T generation.}.
The same video is also uploaded to the repository \href{https://youtu.be/J4MT__Raz_k?si=ZDAQnPCsVrjaSUBv}{YouTube}.

\end{document}


\title{Supplementary Material}

\author{Akihiko~Ikeda}
\email[]{a-ikeda@uec.ac.jp}
\affiliation{Department of Engineering Science, University of Electro-Communications, Chofu, Tokyo 182-8585, Japan}
\author{Yuya~Kubota}
\email[]{kubota@spring8.or.jp}
\affiliation{RIKEN SPring-8 Center, Sayo, Hyogo 679-5148, Japan}
\author{Yuto~Ishii}
\author{Xuguang~Zhou}
\author{Shiyue~Peng}
\author{Hiroaki~Hayashi}
\author{Yasuhiro~H.~Matsuda}
\affiliation{Institute for Solid State Physics, University of Tokyo, Kashiwa, Chiba 277-8581, Japan}
\author{Kosuke~Noda}
\author{Tomoya~Tanaka}
\author{Kotomi~Shimbori}
\author{Kenta~Seki}
\author{Hideaki~Kobayashi}
\author{Dilip~Bhoi}
\affiliation{Department of Engineering Science, University of Electro-Communications, Chofu, Tokyo 182-8585, Japan}
\author{Masaki~Gen}
\affiliation{Institute for Solid State Physics, University of Tokyo, Kashiwa, Chiba 277-8581, Japan}
\affiliation{RIKEN Center for Emergent Matter Science (CEMS), Wako 351-0198, Japan}
\author{Kamini~Gautam}
\affiliation{RIKEN Center for Emergent Matter Science (CEMS), Wako 351-0198, Japan}
\author{Mitsuru~Akaki}
\affiliation{Institute for Material Research, Tohoku University, Sendai, Miyagi 980-0812, Japan}
\author{Shiro~Kawachi}
\affiliation{Graduate School of Science, University of Hyogo, Koto, Hyogo 678-1297, Japan}
\author{Shusuke~Kasamatsu}
\affiliation{Faculty of Science, Yamagata University, Kojirakawa, Yamagata 990-8560, Japan}
\author{Toshihiro~Nomura}
\affiliation{Department of Physics, Faculty of Science, Shizuoka University, Shizuoka 422-8529, Japan}
\author{Yuichi~Inubushi}
\author{Makina~Yabashi}
\affiliation{RIKEN SPring-8 Center, Sayo, Hyogo 679-5148, Japan}
\affiliation{Japan Synchrotron Radiation Research Institute (JASRI), Sayo, Hyogo 679-5198, Japan}
\date{\today}

\maketitle

\section{Analysis of XRD peak profile}
\begin{figure}
\begin{center}
\includegraphics[width = 0.5\columnwidth]{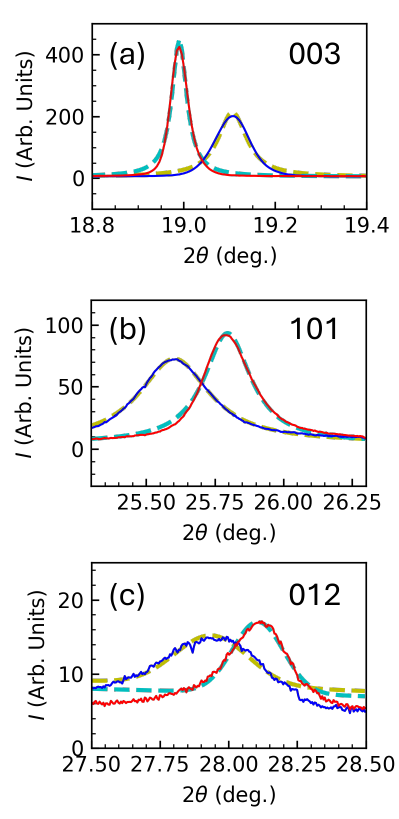}
\caption{
The diffraction peaks of (a) 003, (b) 101, (c) 012 from the power sample of solid \oo at 0 T (red lines) and at 110 T (blue lines) with a temperature of 35 K, which are the same data as those in the Figs. 3(d)-3(f) in the main text.
The Lorentzian fits are shown for 003 and 101. The Gaussian fits are shown for 012.
Yellow, and Magenta lines denote fits to the 110 T data and 0 T data, respectively.
\label{fit1}}
\end{center}
\end{figure}

\begin{figure}
\begin{center}
\includegraphics[width = 0.5\columnwidth]{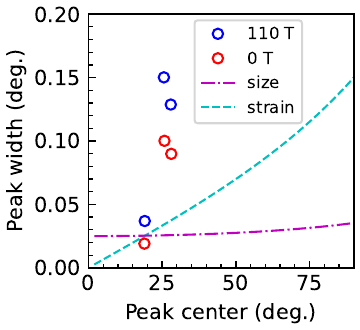}
\caption{
The theoretical curve on the crystal size effect (Dot-dashed line), the microstrain effect (Dashed line), and the experimentally obtained peak width as a function of the peak center in $2\theta$.
\label{fit2}}
\end{center}
\end{figure}

The diffraction profiles are analyzed using the Lorentzian function for 003 and 101, and the Gaussian function for 012, respectively.
The experimental data are the same as those in the Figs. 3(d)-3(f) in the main text.
As shown in Fig. \ref{fit1}, the fits are satisfactory, providing us with fitting parameters of peak positions and peak widths, which are summarized in Table \ref{fittab} and plotted in Fig. \ref{fit2}.

The line width of the diffraction peaks is informative.
It is influenced by factors such as crystalline size, microstress on particles, and dislocations \cite{ZhaoJAC2008, MurrietaMTP2022}.
In Fig. \ref{fit2}, we tentatively compare our results with the particle size effect (Scherrer equation) $\delta_\theta \propto 1 /\cos\theta $ (dot-dashed line) and with the microstrain effect (Stokes-Wilson) $\delta_\theta \propto \tan\theta $ (dashed line).

The analysis shows larger peak widths at scattering angles for 101 and 012 as compared to the lower angle of 003.
It also shows that each peak becomes broader at 110 T as compared to the 0 T data.
The broadening at higher magnetic field may originate in the increase of the microstrain or the decrease of the particle size due to the anisotropic and giant magnetostriction.
However, we are not conclusive on this matter here, because one can see a deviation of the obtained peak width from the common particle size effect or microstrain effect as shown in Fig. \ref{fit2}.
To clarify the origin of the peak width, we require further experimental data collection.

\begin{table}
  \centering
  \caption{Center and width of diffraction peaks}
  \label{fittab}
  \begin{tabular}{lrrrrrr}
    \hline
                &   & 003 &   & 101  &   & 012  \\
                & 0 T &  110 T &  0 T &  110 T & 0 T &  110 T \\
    \hline
    Center (deg.) & 18.99 & 19.11 & 25.80 & 25.60 & 28.11 & 27.94 \\
    Width (deg.)    & 0.0194 & 0.0374 & 0.1027 & 0.1541 & 0.0926 & 0.1327 \\
    \hline
  \end{tabular}
\end{table}

\bibliography{pink02}